\documentclass{iaus}
\usepackage{graphicx}
\usepackage{dcolumn}
\usepackage{amsmath}
\usepackage{epsfig}
\usepackage{bm}
\usepackage{amssymb}

\title[IAUS266.~~Star Clusters and dSph galaxies] %% give here short title %%
{Star clusters as building blocks for \\ dSph galaxies formation}

\author[P. Assmann et al.]   %% give here short author list %%
{P. Assmann$^1$,
  M. Fellhauer$^1$ \and M. I. Wilkinson$^2$}

\affiliation{$^1$Departamento de Astronom\'{\i}a, Universidad de Concepci\'{o}n, \\ Casilla 160-C,
Concepci\'{o}n, Chile \\ email: {\tt passmann@astro-udec.cl} \\ email: {\tt mfellhauer@astro-udec.cl} \\[\affilskip]
$^2$Dept. of Physics \& Astronomy, University of Leicester, \\ Leicester, UK \\email: {\tt miw6@astro.le.ac.uk}}

\pubyear{2009}
\volume{266}  %% insert here IAU Symposium No.
\pagerange{119--123}
\jname{Star Clusters: Basic Building blocks \\ Throughout Time And Space}
\editors{Richard de Grijs , Jacques Lepine, eds.}
\begin{document}

\maketitle

\begin{abstract}
We study numerically the formation of dSph galaxies. Intense star bursts, e.g. in gas-rich environments, typically produce a few to a few hundred young star clusters, within a region of just a few hundred pc. The dynamical evolution of these star clusters may explain the formation of the luminous component of dwarf spheroidal galaxies (dSph). Here we perform a numerical experiment to show that the evolution of star clusters complexes in dark matter haloes can explain the formation of the luminous components of dSph galaxies.
%% add here a maximum of 10 keywords, to be taken form the file <Keywords.txt>
\end{abstract}

\firstsection % if your document starts with a section,
              % remove some space above using this command.
\section{Introduction}
Understanding the formation and evolution of galaxies is a fundamental goal in astronomy. Galaxies are believed (in the standard cosmological scenario) to build up by the merging of smaller constituents, namely dwarf galaxies. This scenario makes dwarf galaxies the basic building blocks of the structure formation in the universe (e.g. \cite[Springel et al. 2005]{springel05}). Therefore, to understand galaxy formation and evolution on all scales it is crucial to understand the formation and evolution of the smallest constituents first. Dwarf galaxies come in different shapes and dynamical states, e.g. dwarf disc galaxies, dwarf elliptical galaxies (with or without nucleus), dwarf irregulars, ultra-compact dwarfs and dwarf spheroidal (dSph) to name a few. According to the standard scenario all galaxies reside in an extended dark matter (DM) halo, whereby dwarf galaxies should reside in the highest M/L DM haloes (relative to their luminous, baryonic mass) of all galaxies.

 We investigate numerically the initial conditions for the formation of dSph galaxies. Our model is based on the assumption that stars form in star bursts (\cite[Lada \& Lada 2003]{lada03}), producing star clusters, within a region of just a few hundred pc. The dynamical evolution of these star clusters, i.e their dissolution due to gas-expulsion and subsequent merging in the center of the DM halo, may explain the formation of dSph galaxies. Our numerical simulations are carried out using the particle mesh code Superbox (\cite[Fellhauer et al. 2000]{fell00}), which allows us to keep track of many objects in one simulation. The distribution of the star clusters inside the central area of the DM halo follows a Plummer distribution, while we take two distinct types of DM haloes into account: a cusped Navarro-Frenk \& White (NFW)(\cite[Navarro et al. 1997]{nav97}) and a cored Plummer profile (\cite[Plummer, H. C. 1911]{plum11}). The setup of our simulation is as follow: the single star clusters are represented by Plummer spheres with a Plummer radius of $4$~pc and a cutoff radius of $25$~pc. Each cluster has an initial mass of $10^{5}$~$M_{\bigodot}$ and an initial crossing time of $2.4$~Myr, and is represented with 100,000 particles. Fifteen of these clusters are placed in a star cluster (SC) complex that is modeled again as a Plummer distribution (i.e., positions and velocities according to the Plummer distribution function), now with the star clusters as particles. This Plummer distribution is given a Plummer radius of 25 pc, a cutoff radius of 150 pc. The dark matter haloes have a scale-length of 500 pc and the mass enclosed within one scale-length of $1.01\times10^{8}$ $M_{\bigodot}$, using $1,000,000$ particles. For the dark matter halo NFW profile (Plummer profile), we used a cutoff radius of $16$~kpc ($2.5$~kpc).

\section{Experimental Results}

\begin{figure}[h]
 %\vspace*{0.5 cm}
\begin{center}
 \centerline{\includegraphics[width=5.4in]{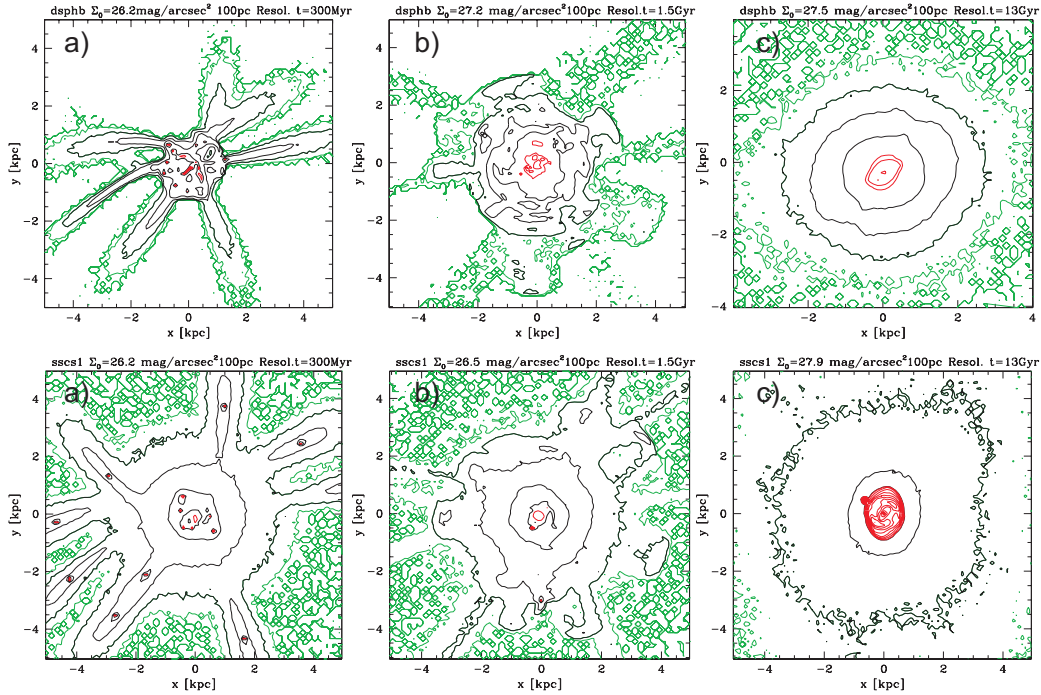}}
 %\vspace*{0.1 cm}
 \caption{Top: dSph fomation based on SCs dissolution in a cored DM halo. The surface density is shown at a) 300 Myr, b) 1.5 Gyr c) 13 Gyr.
 Bottom: dSph fomation based on SCs dissolution in a cusped DM halo. The surface density is shown at a) 300~Myr, b) 1.5~Gyr c) 13~Gyr.  }
   \label{Fig:plum}
\end{center}
\end{figure}

In Fig.~\ref{Fig:plum} top (bottom) , we show a time sequence of surface density contours of the star clusters traveling through the central area of the cored (cusped) DM halo, while expanding due to gas expulsion, collisions, etc. This leads to a subsequent dissolution of the SCs and the distribution of their stars in the central area of the DM halo, building up the luminous component of the dSph galaxy.

\begin{figure}[h]
 %\vspace*{0.5 cm}
\begin{center}
 \centerline{\includegraphics[width=5.4in]{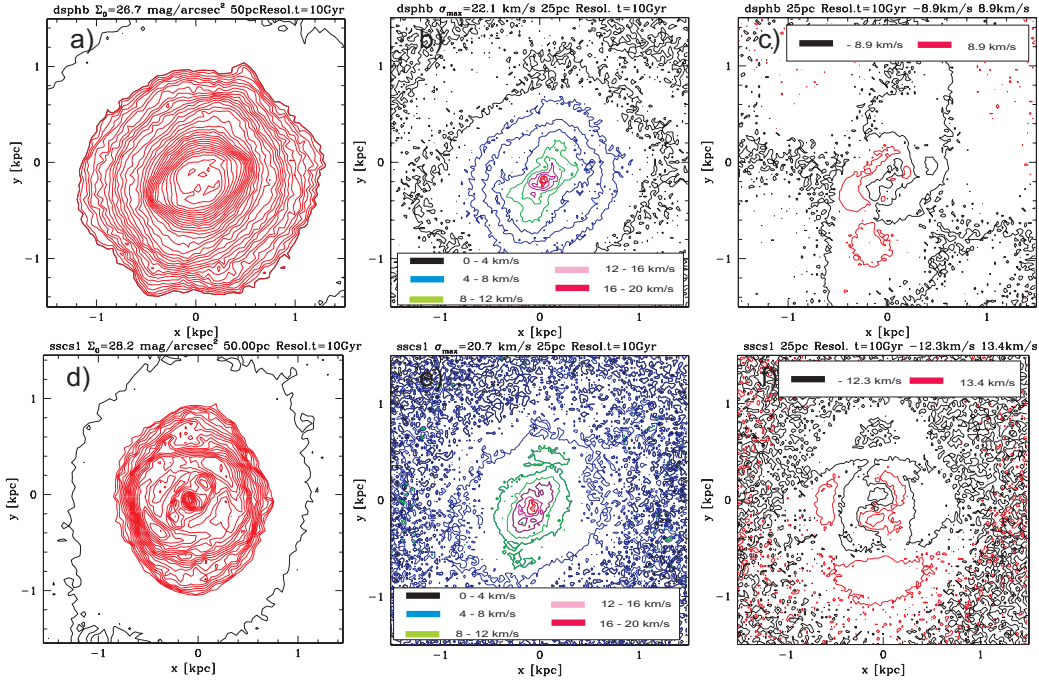}}
 %\vspace*{-0.5 cm}
 \caption{Projection of the dSph onto the xy plane. In a) and d) the surface density with 50 pc resolution for the Plummer and NFW profile, is shown; In b) [e)] the velocity dispersion with 25 pc resolution for the Plummer profile [NFW profile] is shown. In c) and e) we show the mean velocities of the stars for the Plummer and NFW profile.}
   \label{Fig:zoom}
\end{center}
\end{figure}

In Fig.~\ref{Fig:zoom} for both scenarios of DM haloes, the outer contours of our objects are rather smooth as have been observed in most of the larger dwarf galaxies close to the Milky Way. However, it is interesting to note that there are still some fluctuations visible in the central part, and observations of this fluctuation might corroborate our model for dSph galaxy formation. We note that in the cusped simulation, the stars are more centrally concentrated.  In the middle panels we can see that the velocity dispersion is still about $50 \%$ higher than expected considering the observations of Draco or Ursa Minor dSph, but focusing at the core of our simulated objects, in the cusped simulation the dispersion velocities are lower than the cored simulation. The mean velocities "per pixel" shown in the right panels, shows areas which differ by ± 9 km/s in the cored simulation and ± 13 km/s for the cusped simulation from the average which should be zero. We call these "streams" fossil remnants of the formation process. With the advent of new techniques for astronomical observations, these streams should be detectable and can be used to test our formation scenario.

\begin{figure}[h]
 %\vspace*{0.5 cm}
\begin{center}
 \centerline{\includegraphics[width=3.8in]{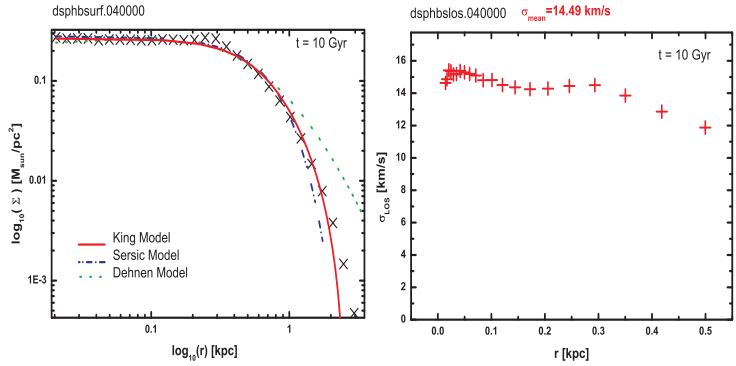}}
 %\vspace*{0.5 cm}
 \caption{a) The surface density distribution fitted with different profiles. b) The line of sight velocity dispersion profile.}
   \label{Fig:king}
\end{center}
\end{figure}

The surface brightness profile for the cored simulation after $10$~Gyr of evolution, is shown in Fig.~\ref{Fig:king}(left panel). We fit our model with King, Sersic and Dehnen profiles. Our results are best described  by the King profile. There the core radius is $720$~pc and the central surface brightness is $26$~$mag/arcsec^{2}$ . In the right panel in Fig.~\ref{Fig:king} we show the line of sight velocity dispersion within a radius of $500$~pc. The central velocity dispersion is $14.5$~km/s.

\begin{figure}[h]
 %\vspace*{0.5 cm}
\begin{center}
\centerline{\includegraphics[width=1.5in,height=1.5in]{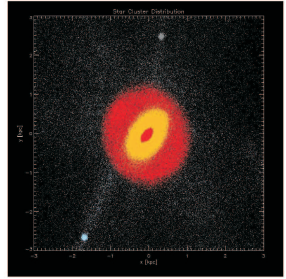}}
 %\vspace*{0.5 cm}
 \caption{The red and yellow points represent two types of streams of stars at 10 Gyr in the central region of the dSph. We can also see two small star clusters orbiting the galaxy.}
   \label{Fig:star}
\end{center}
\end{figure}

In Fig.~\ref{Fig:star} we show some stars clusters after$10$~Gyr of evolution for the cusped simulation. The two SC shown represent the two main types of stellar streams we see. One from a radial orbit has spread its stars uniformly in a spherical region, while the other one, on a more circular orbit, has stars in a torus-like configuration. These torus-like "streams" are responsible for an artificially raised velocity dispersion in projection. One also sees two surviving star clusters in the outskirts of the galaxy similar to the SC in Sagittarius and Fornax.

\section{Conclusion \& Outlook}

We study numerically the formation of dSph galaxies out of merging SC, considering two distinct types of profiles for the dark matter halo. We compare the results of both simulations. We see that in the cusped simulation, stars are more centrally concentrated but have lower central velocity dispersions. In both simulations there are streams of stars (fossil remnants of the dissolved SCs) which survive more than 10 Gyr of evolution,
raising the velocity dispersion artificially. With the advent of new techniques for astronomical observations, these streams should soon be detectable and test our formation scenario. Our initial setup was chosen to produce a stellar component with a line-of-sight velocity dispersion of 10~km/s, similar to the larger dSph galaxies of the Milky Way. But due to the stellar "streams" we see an artificially raised dispersion in both models. These streams can only produce such effect in a low density environments like dSph galaxy.

We will investigate this effect further and if we find that is persistent substructure, we will explore the potential impact of undetected streams on estimates of the DM content of the MW dSph galaxies (based on measurements of the velocity dispersion).

\end{document}